# Damping and vibration properties of alginate-poloxamer hydrogels doped with sepiolite and cactus fibres

Gianni Comandini[a], Fabrizio Scarpa[a,1], Evita Ning[b], Graham J Day[a,c], James P K Armstrong[d], Abderrezak Bezazi[e], Adam W Perriman[b,f]

[a]Bristol Composites Institute, School of Civil, Aerospace and Design Engineering (CADE), University of Bristol, Bristol BS8 1TR, United Kingdom
[b]School of Cellular and Molecular Medicine, University of Bristol, Bristol BS8 1TD, United Kingdom
[c]Centre for the Cellular Microenvironment, Division of Biomedical Engineering, James Watt School of Engineering, The Advance Research Centre, University of Glasgow, Glasgow G12 8QQ, United Kingdom
[d]Translational Health Sciences, Bristol Medical School, University of Bristol, Bristol BS1 3NY, United Kingdom
[e]Laboratoire de Mécanique Appliquée des Nouveaux Matériaux, Université 8 Mai 1945, Guelma, Algeria
[f]Research School of Chemistry and John Curtin School of Medical Research, Australian National University, Canberra ACT2601, Australia

## Abstract
We investigated a novel class of composite hydrogels composed of alginate, poloxamer, sepiolite, and cactus fibres for vibration damping applications. Using a Design of Experiments methodology, we systematically correlated manufacturing parameters with mechanical and damping properties, performance using Dynamic Mechanical Analysis and vibration testing. The hydrogels were characterised under controlled temperature, frequency, and humidity conditions, with results demonstrating that the storage modulus can reach up to twice that of pure hydrogel formulations, using diluted dispersions with total additive concentration below 2 wt%. Sepiolite additions below 0.3 wt% were found to stabilise the material response to temperature variations, while cactus fibres enhanced both stiffness and damping performance in a concentration-dependent manner. Optimal performance was achieved with a formulation containing of 5 wt% alginate, 5 wt% poloxamer, 0.1 wt% sepiolite, and 2 wt% cactus fibres. This composition provided a favourable balance between quasi-static mechanical integrity and dynamic damping capability, with loss factors exceeding 0.4. These findings provide a foundation for developing advanced and sustainable hydrogel materials with tailorable vibration damping characteristics.

---
[1] Corresponding author: f.scarpa@bristol.ac.uk

# 1. Introduction

Vibration and acoustic damping technologies have a long history, with references and device designs for vibration mitigation found in the works of Lord Rayleigh[1], Bishop[2] and Kolsky[3]. Vibration and damping treatments were also developed in the early 1940s by both Allied and Axes forces[4,5]. However, the designs of passive vibration, thermal, and acoustic insulators have changed little since the introduction of multi-layer insulators and polymeric constrained/free layer dampers in the early 1960s, which are used to stabilise cooling systems and manage flutter-type and acoustically radiating vibrations in aircraft, helicopters, and marine structures. Damping and energy dissipation performance of free and constrained layers dampers depend on their architecture and on the viscoelasticity of their polymers, with varying storage and damping (tanδ) based on temperature-frequency equivalence via Arrhenius or Williams-Landel-Ferry relations[6]. Typical elastomers used for these damping technologies have moduli varying between 5 MPa and 2.6 GPa when passing from a rubbery to glass state. Soft polymers that are mostly used in constrained layer dampers have moduli varying between 170 kPa and 100 MPa between 90°C and -37°C, with damping (loss) factors ~ 0.9 at 25°C[4]. When used within vibrating composite beams, the overall loss factor of the system however drops to ~ 0.3[7]. At a constant temperature, the higher the excitation frequency provided by the dynamic load, the more the polymer approaches the glassy state and becomes dynamically rigid, resulting in a loss factor decrease.

Most of the materials used in current state-of-the-art damping and absorption technologies are based on manipulation of synthetic chemistry and are fundamentally fossil-like in nature. Biobased materials, however, provide an alternative to fossil-based solutions. Natural fibres (flax, hemp, wood) have tanδ/damping loss factors between 8% and 20% when embedded within thermoplastics and thermoset composites (PLA and epoxies), with the relative humidity increasing the damping by 50% when passing from 20% to 80% of relative humidity[8]. Hydrogels[9] can also be produced using biobased and biodegradable sources and offer some interesting behaviour in terms of damping capacity. Recoverable energy dissipation from drop tower impact tests has been observed in sodium alginate-silicon nitride PVA composite biomaterials[10]. Lead zirconate titanate (PZT)-polydimethylsiloxane (PDMS) gels. have shown loss factors of ~20% within 1–20 Hz sweeps performed with dynamic mechanical analysers[11]. Wang *et al.* analysed polyacrylamide (PAAm) and PDMS gels using a bending rig with base acceleration. PAAm gels had viscous damping ratios of ~2% to 3%, while PDMS beams showed ratios around 8%[12]. Polyurethane-carbon nanotube-PDMS gel composites have shown loss factors exceeding 24% in vibration transmissibility tests between 50 and 600 Hz[13]. Hydrogels based on PVA and alginate/poloxamer can also be used as a platform to host reinforcing natural fibres, such as flax[14]. Diluted dispersions of flax fibres provide increased stiffness and strength, higher water absorption capability but lower thermal stability compared to other natural fibre reinforcements like jute. Another class of natural fibre reinforcements is represented by cactus. Cactus fibres are derived from sheaths of *Opiuntus ficus Indica* (prickly pear). The plant is native and endemic to the Mediterranean region, North Africa, Latin America, Australia, and Texas, where it is commonly used as feed for cattle[15]. The fibres have shown to exhibit multiscale fractal geometry characteristics that enhance stick-slip friction and bonding to the matrix [16-18]. Slip-stick friction effects have also been demonstrated from preliminary vibration transmissibility tests on alginate/poloxamer gels up to 5.0 wt% [19]. However, nanoclays constitute a more widespread type of reinforcement in hydrogel composites. Sepiolite ($Si_{12}O_{30}Mg_8(OH)_4(H_2O)_4 \cdot H_2O$) is a particularly interesting type of nanoclay formed by precipitation of near-surface brackish or saline waters in semi-arid climatic conditions[20]. The special structure of the sepiolite generates nanofillers with large aspect ratio and high surface area for significant sorption capability. In addition, the negative surface charge of sepiolite allows strong complexation with positively charged polymers or surfaces. These properties have been used to develop layer-by-layer coatings with polymers like chitosan and polyacrylic acid (PAA) to enhance the vibration damping, dynamic stiffness and impact energy absorption in foams and other porous materials used for packaging[21,22]. Sepiolite is considered one of the most efficient nanoclay fillers to produce environmentally sustainable hydrogel systems for retaining water and dyes[23], antibacterial and self-healing properties[24] and significant increase of storage moduli and loss factors observed during shear rheology tests[25,26].

Recent work[19,27] has demonstrated that alginate/poloxamer hydrogels form tunable porosity and can achieve dynamic moduli over ~ 2.5 MPa and loss factors between ~15% and 28% within the 80 Hz - 300Hz frequency range under vibration transmissibility tests. Bioprinted cactus microfibre-doped alginate/poloxamer hydrogel monoliths[18] when subjected to vibration excitation also tend to increase the dynamic modulus of almost an order of magnitude compared to their static moduli, and loss factors between ~18%-30%. The cactus provides a stiffening effect at low concentrations, however the synergy between slip-stick effects provided by the fractal cactus fibres[16] and the alginate/poloxamer networks generates a ~ one order of magnitude increase of the dynamic modulus between 50 Hz and 150 Hz under vibration[19]. To provide an indication of the significance of these numbers, state-of-the-art polyurethane foams used for energy absorption/packing show a maximum increase in the dynamic modulus of 50% when compared to the static modulus of ~100 kPa due to the stiffening poroelastic effect provided by the gas trapped within the tortuous foam skeleton[28]. However – to the best of our knowledge – no data are available about the viscoelasticity and vibration damping performance of alginate/poloxamer hydrogel composites containing sepiolite reinforcements, or combined sepiolite/natural fibre cactus fillers. Overall, there is a significant gap in our understanding of how various types of nanoclays and natural fibre reinforcements contribute to the quasi-static and dynamic performance of biobased and sustainable hydrogels, such as alginate/poloxamer.

This work describes a systematic design of experiments (DoE) campaign to assess the quasi-static, viscoelastic and vibration transmissibility properties of novel alginate/poloxamer hydrogels reinforced by sepiolite and cactus fibre reinforcements. A fractional factorial design involving 53 different experiments has been carried out to assess storage and loss moduli with dynamic mechanical analysis (DMA), and loss factors and dynamic modulus via vibration transmissibility tests. Significantly, the DMA tests are carried out at constant relative humidity levels (20% RH) to avoid effects from lack/acquisition of moisture from the environment.

## 2. Design of Experiments (DoE) and Materials

The materials involved in this Design of Experiments are poloxamer, alginate, sepiolite and cactus fibres. The alginate/poloxamer base of the hydrogels was kept constant at 5 wt% alginate and 5 wt% poloxamer, following the optimal baseline composition identified in[19]. A Design of Experiments was implemented following a fractional factorial design approach[29]. The cactus fibres used in this work have been ball-milled with a cryo mill (zirconium balls); the resulting fibres had average thickness dimensions between 2.5 and 3 μm. Poloxamer 407 was incorporated into the hydrogel formulation to regulate the gelation temperature, improving both injectability and the practical usability of the material. Maintaining a constant concentration of 5 wt% across all samples standardised its thermoresponsive behaviour, ensuring consistent influence on gelation and mechanical properties. Sodium alginate was also included in the formulation as a constant component, maintained at a 5 wt%[19]. The 5 wt% alginate/poloxamer was also previously identified as the hydrogel configuration offering the best trade-off in terms of quasi-static mechanical properties and vibration damping[30]. Weight fractions between 0.1 and 0.5% of sepiolite have previously shown to provide significant increase of loss modulus in chitosan[21] and poly(acrylic) acid (PAA)-based[22] nanocoatings deposited on open cell polyurethane foams. Weight fractions of cactus fibres between 0.5 and 2% have also shown in the previous generation of alginate/poloxamer/cactus fibre composites the best trade off in terms stiffness and energy absorption[19].
The fractional factorial design resulted in a total of 16 replicated experimental conditions for a total of 53 experiments (Table 1).

*Table 1. Summary of hydrogel samples prepared with varying weight percentages of sodium alginate, Poloxamer 407, sepiolite and cactus fibres.*

| Sample type | Sodium alginate (wt%) | Poloxamer 407 (wt%) | Sepiolite (wt%) | Cactus fibres (wt%) |
|---|---|---|---|---|
| 1 | 5% | 5% | 0% | 0% |
| 2 | 5% | 5% | 0.1% | 0% |
| 3 | 5% | 5% | 0.3% | 0% |
| 4 | 5% | 5% | 0.5% | 0% |
| 5 | 5% | 5% | 0.1% | 0.5% |
| 6 | 5% | 5% | 0.1% | 1% |
| 7 | 5% | 5% | 0.1% | 2% |
| 8 | 5% | 5% | 0.3% | 0.5% |
| 9 | 5% | 5% | 0.3% | 1% |
| 10 | 5% | 5% | 0.3% | 2% |
| 11 | 5% | 5% | 0.5% | 0.5% |
| 12 | 5% | 5% | 0.5% | 1% |
| 13 | 5% | 5% | 0.5% | 2% |
| 14 | 5% | 5% | 0% | 0.5% |
| 15 | 5% | 5% | 0% | 1% |
| 16 | 5% | 5% | 0% | 2% |

A 12.5% w/w Poloxamer 407 solution was prepared by gradually dissolving 2.5 g of Poloxamer 407 into 20 g of precooled deionised water (4 °C) under continuous stirring, maintaining the temperature throughout. Mixing continued until a homogenous, clear solution was obtained. Poloxamer 407 is thermo-responsive, remaining in a liquid state below 10–15 °C and undergoing gelation above 30–37 °C. Gel formation was induced by incubating the solution at 37 °C. Separately, a 12.5% w/w sodium alginate solution was prepared by dissolving 2.5 g of sodium alginate in 20 g of deionised water. The two solutions were subsequently combined to yield a uniform composite mixture. To achieve a final target mass of 50 g, an additional 10 g of deionised water (dH$_2$O) was initially intended for inclusion. However, when sepiolite and/or cactus fibres were incorporated as additives, their respective masses were subtracted from this 10 g allocation to maintain a constant total mass. The resulting mixtures were subjected to three mixing cycles at 2500 rpm for 5 minutes each, with 15 min rest intervals between cycles, ensuring thorough homogenisation and degassing. The blended formulations were then cast into moulds and enclosed within dialysis membranes. These encapsulated samples were immersed in a 2 L bath of dH$_2$O containing 147.01 g of calcium chloride (500 mM) at 37 °C for 24 h to induce Ca$^{2+}$ mediated alginate cross-linking. Following gelation, the hydrogels were removed from the dialysis membranes and stored in 1 L of dH$_2$O supplemented with 14.7 g of calcium chloride (100 mM) to preserve structural integrity prior to further testing.

## 3. Testing Methods

Microstructures of the gels were analysed using cryo-scanning electron microscopy (Quanta 200 with Field Emission Gun and Gatan Alto 2500 Cryo-SEM attachment). The viscoelastic properties of the composite hydrogels were evaluated using a Dynamic Mechanical Analysis facility (DMA 850 from Texas Instruments). Tests were performed in relative humidity (RH) chamber maintained at constant 20% RH, a value substantially lower than the standard ambient level of approximately 50% at standard room temperatures. The samples were mounted using a submersion clamp and subjected to cyclic compression loading at 1 Hz, a frequency commonly used to evaluate the viscoelastic response of hydrogels[10,31]. The temperature was varied between 3°C to 95°C; however, only data up to 60°C were retained for analysis. This temperature range was selected to evaluate the thermal stability and performance of the hydrogels under conditions exceeding physiological incubation temperature. Notably, the gels exhibited significant compression and dimensional reduction (by a factor of 3) after repeated harmonic cycling, especially at temperatures above 70°C. This deformation results in a stiffening response, manifested as an increase in the modulus measured. For this purpose, only data up to a maximum compressive strain of 10% were considered to extract properties close to the linear elastic regime in viscoelasticity for hyperelastic polymers[32].

Vibration transmissibility tests have been carried out to determine the dynamic modulus and loss factors[28] of the hydrogels within frequency ranges from 80 Hz to 180 Hz at room temperature. The specimens used for the vibration transmissibility tests were dimensioned at 30 mm × 30 mm × 15 mm. The experimental protocol involved evaluating these specimens under four distinct top mass conditions: 74.9 g, 121.6 g, 168.6 g, and 239.2 g. These masses were affixed to the upper plate, while the hydrogel samples were affixed to the upper and lower plates. A DEHP Alginate Tray Adhesive Liquid was used to ensure secure contact between the plates and the sample. The lower plate was connected to an electrodynamic shaker (LDS, Model V406), which delivered controlled vibrational excitation. To quantify the vibration transmission, two accelerometers (PCB, Model 333M07) were mounted to measure acceleration at both the base and upper plates. Each test consisted of two cycles, with each cycle consisting of five 10-second segments. During each segment, white noise was applied at varying energy levels, yielding a total test excittion time of 50 seconds per cycle. A 10-second interval between cycles was introduced to allow for system stabilisation. The Root Mean Square (RMS) acceleration applied to the base plate was controlled and set to values of 4.6, 7.6, 12.0, 17.9, and 22.5 m/s².

## 3. Results

### 3.1 Microstructure of the composite hydrogels

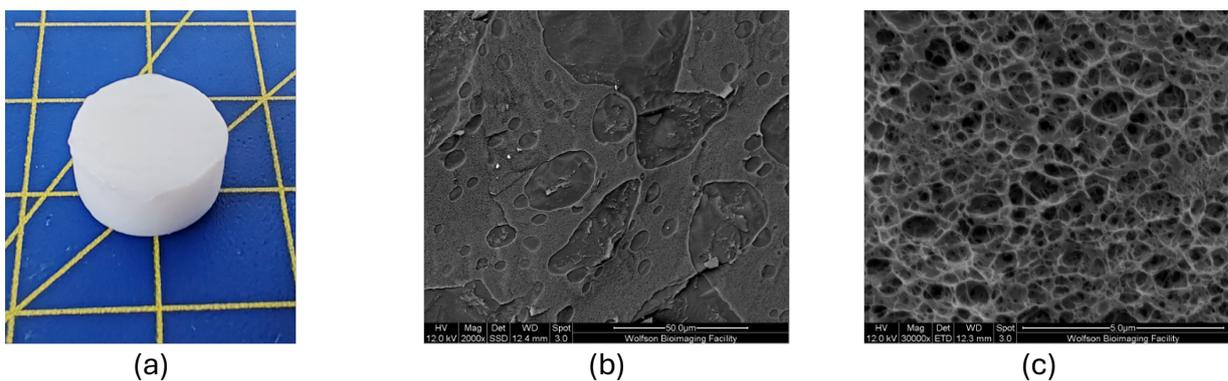

(a)              (b)              (c)

*Figure i. (a) Alginate/Poloxamer 5 wt% cross-linked with $CaCl_2$, prepared f for DMA testing within an environmental chamber. Cryo-SEM images of the microstructure at different magnifications (b-c), illustrating the porogen-based architecture of the hydrogels.*

The baseline alginate/poloxamer hydrogel showed the presence of a structured 3D network, with pore diameters ranging from approximately 1.3-15 μm (Figure i). Those dimensions are consistent with the porogenic effect induced using Poloxamer[27,30]. The incorporation of the sepiolite and cactus fibre reinforcements alters the microstructure of the composite hydrogels (Figure ii). Both sepiolite and cactus fibres appear to form good interfacial bonding with the alginate/poloxamer matrix. The rough surface texture of the cactus fibre (Figure iif) enhances the bonding to the matrix and the load transfer during mechanical loading[17].

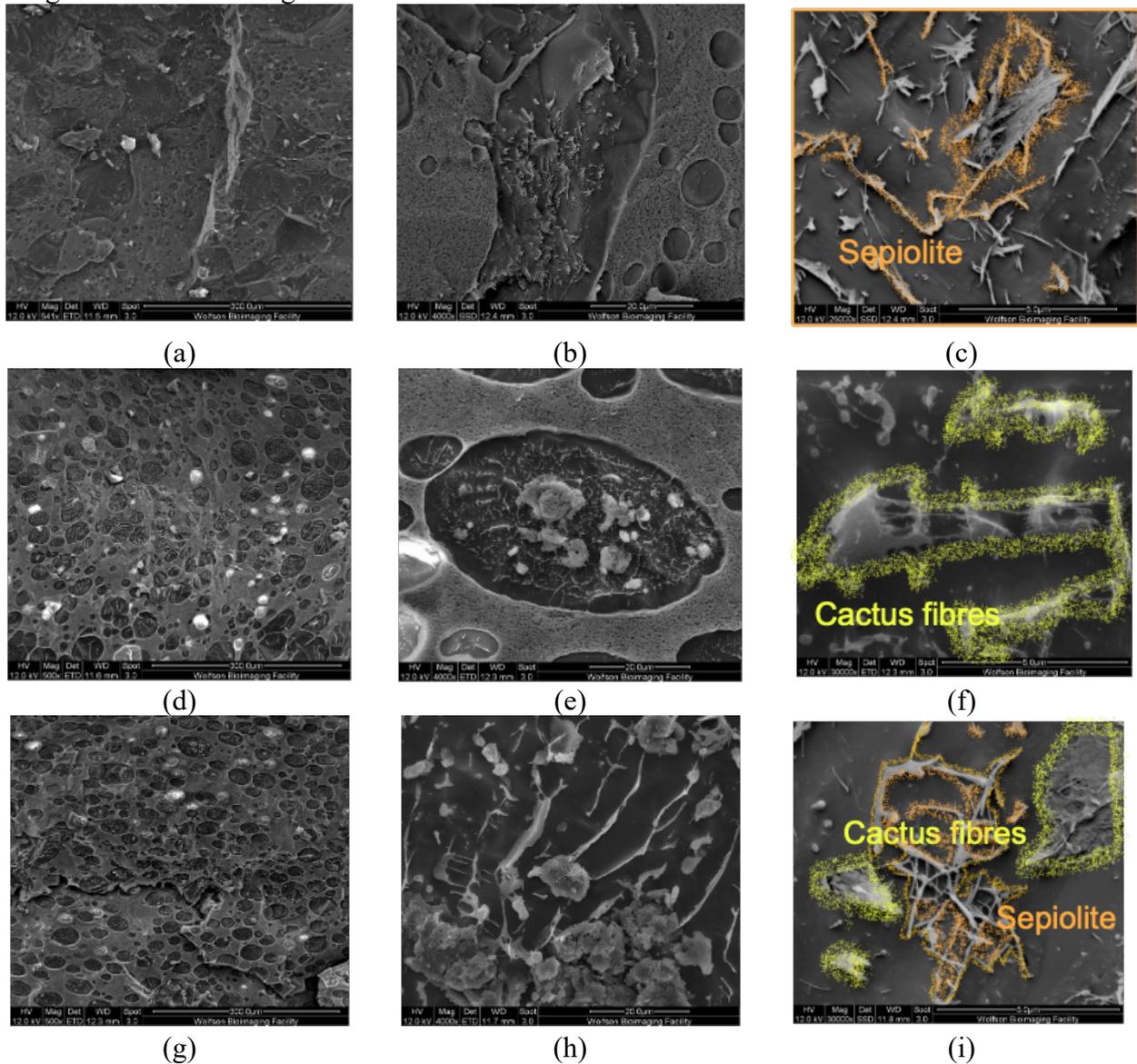

Figure ii. Cryo-SEM images of the microstructures of the composite hydrogels at 300 $\mu m$ (a, d, g), 20 $\mu m$ (b, e, h) and 5 $\mu m$ (c, f, i) details of magnification. Images (a-c) are related to composites with 0.3 wt% of sepiolite reinforcement, (d-f) with cactus reinforcement only at 2 wt%, and (g-i) to samples with sepiolite 0.1 wt% and cactus at 1 wt%. Cactus fibres and sepiolite reinforcements are artificially contoured in green and orange, respectively.

## 3.2 Quasi-static DMA tests

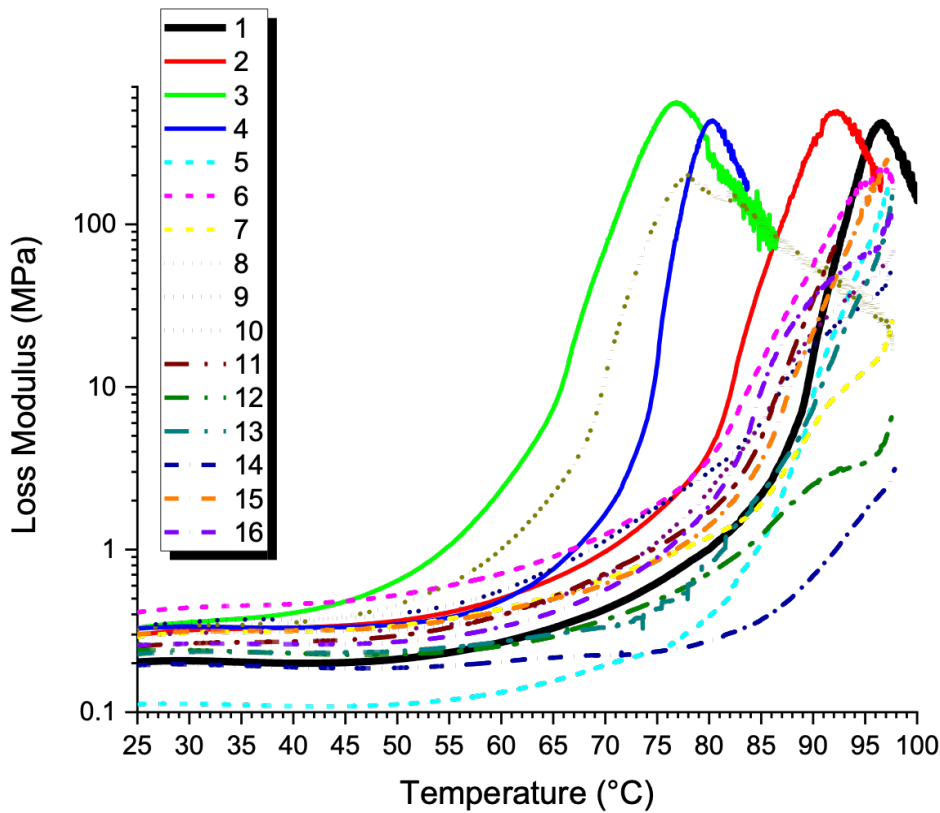

*Figure iii. Example of loss modulus curves between 25°C and 100°C of all the samples tested with the DMA at 1 Hz and 20% RH.*

Figure iii displays the overall loss modulus curves versus temperature for a subset of the gels tested within the Design of Experiments. Peaks of loss modulus for the different hydrogel composites occurred between approximately 75°C and 95°C. Loss moduli tended to feature a low sensitivity versus the temperature within the ~25°C – 60°C range, with a steep increases up to ~ 3 orders of magnitude when approaching the peak of the loss modulus. A more detailed comparison at the distribution of the storage and loss moduli for the different combinations of hydrogels can be observed in Figure iv at 30°C, 45°C and 60°C. The baseline alginate/poloxamer gels exhibited stable values of storage and loss moduli across these three different temperatures (~1.2 MPa and ~ 0.22 MPa, respectively). The deviations however increased with the increasing testing temperature. Notably, alginate/poloxamer gels are primarily designed for tissue engineering applications at – 37°C, where their structural integrity and material homogeneity are optimal. The energy absorption (loss modulus) of the 0.3 wt% sepiolite was largest at 60°C (0.7 MPa), although variability was substantial with a high deviation. This likely reflects reduced structural integrity at elevated temperature, which compromises the microstructure of the undoped gel. The gels with 0.1 wt% sepiolite and 1 wt% cactus also exhibited considerable loss moduli at 30°C and 45°C (~0.37 MPa - 0.6 MPa). While the standard deviations remain high at 60°C, they were lower than 0.3 wt% sepiolite samples. Notably, 0.3 wt% sepiolite and 1 wt% cactus fibre formulations showed relatively low standard deviations and consistent energy absorption values ranging from 0.3 MPa at 30°C to 0.55 MPa at 60°C.

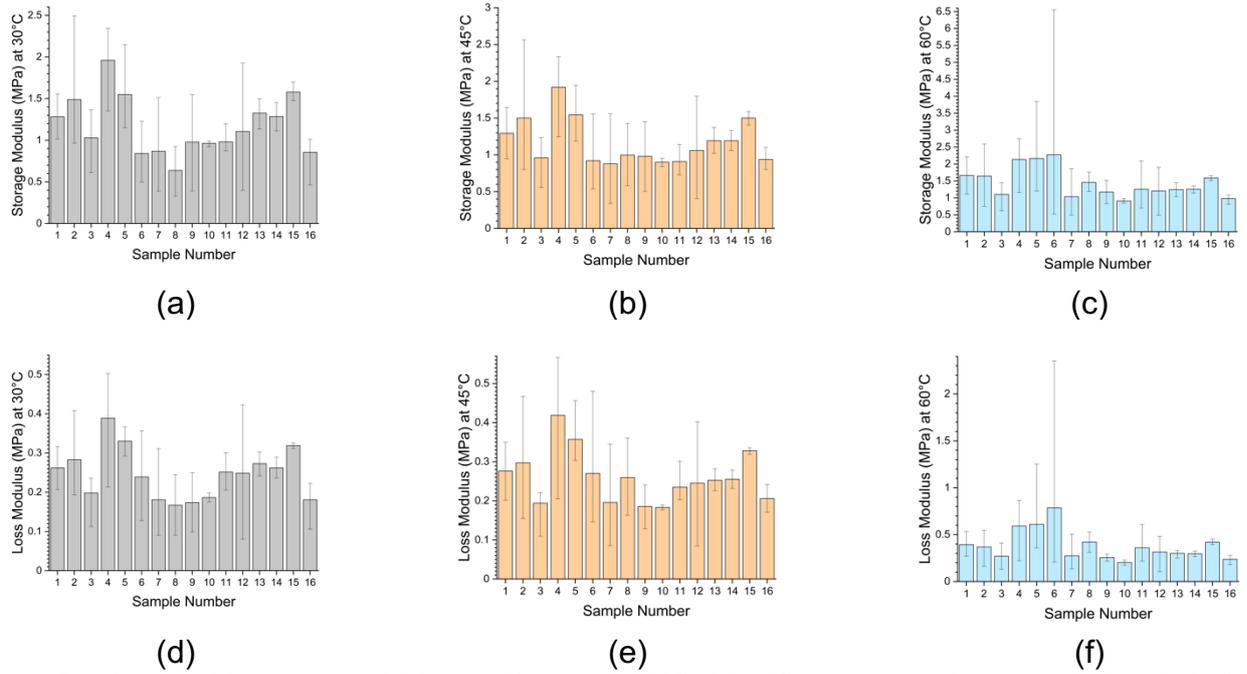

*Figure iv. Distributions of the storage moduli (a-c) and loss. moduli (d-f) of the different classes of composite hydrogels. (a,d) are referred to values measured at 30°C, (b, e) at 45°c and (c, f) at 60°C.*

### 3.3. Vibration transmissibility tests

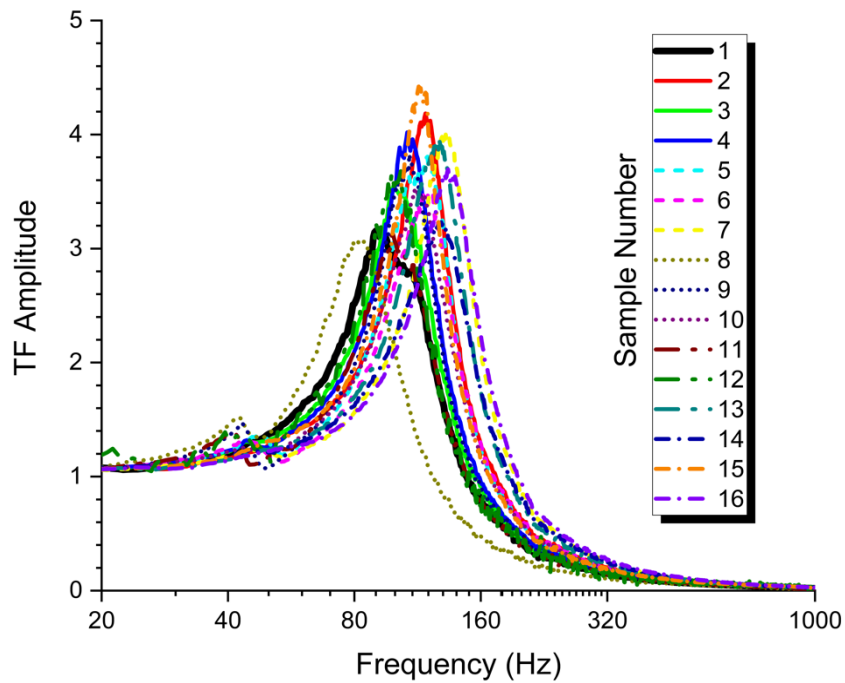

*Figure v. Distribution of the transmissibility curves for the different types of composite hydrogel samples measured at room temperature.*

Figure v shows the cumulative transfer function plots from the vibration transmissibility tests conducted on various gel systems. The resonance frequencies are generally clustered between 80 Hz and 140 Hz. Minor peaks in the transfer function are observed around 40 Hz; however, these are attributed to slight rocking motions of the gel/top mass system, likely caused by minor eccentricities

in the dynamic load. The very small amplitude of this rocking resonance does not appear to affect the primary uniaxial/compressional dynamic characteristics measured during the vibration transmissibility tests.

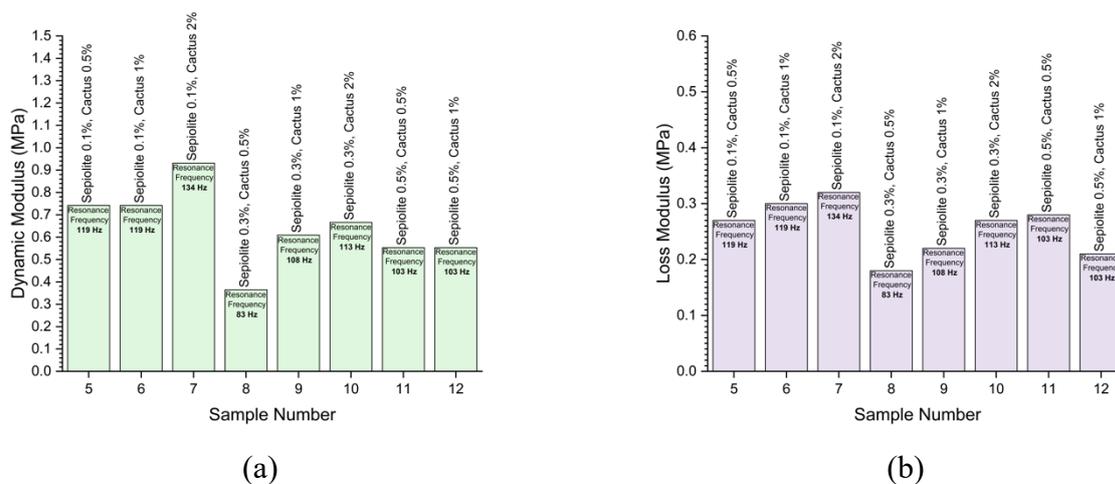

*Figure vi. Maps of (a) dynamic moduli and (b) loss factors measured with the vibration transmissibility tests.*

The distributions of dynamic moduli, loss factors, and resonant frequencies for the different gels' formulations are shown in Figure vi. Within the 80 Hz and 150 Hz range the highest loss modulus values were observed in gels containing 0.1 wt% sepiolite and up to 2.0 wt% cactus, reaching 0.32 MPa. The corresponding dynamic modulus for this configuration is 0.93 MPa. Notably, gels with 0.5 wt% and 1.0 wt% cactus fibres exhibited similar loss modulus values around 0.3 MPa, indicating a plateau in energy dissipation performance at these intermediate concentrations. Increasing the sepiolite concentration beyond 0.1 wt% does not result in higher dynamic or loss modulus values, instead a reduction of 10% to 20% was observed compared to the 0.1 wt% sepiolite formulation. However, increasing the cactus fibre content appeared to enhance performance at a constant weight fraction of sepiolite. For instance, in gels with 0.3 wt% sepiolite, the loss modulus increased monotonically from 0.18 MPa with cactus at 0.5 wt%, to 0.27 MPa at 2.0 wt%. This positive trend with increasing cactus fibre reinforcement was consistent across all sepiolite weight fractions investigated.

## 4. Discussion

### 4.1 Role of sepiolite

Previous studies using mechanical testing and shear rheology have demonstrated that incorporating 10 wt% of sepiolite into polyvinyl alcohol (PVA) composite gels can increase toughness by a factor of 60 and enhance the loss modulus at 10 Hz and 25°C by 50%, respectively[25]. Higher sepiolite concentrations (20 wt%-40 wt%) in castor-oil gel-like dispersions have similarly resulted a 5000-fold increase in the loss modulus[26]. Furthermore, sepiolite has been shown to maintain thermal and chemical stability in paraffin composites subjected to over 200 thermal cycles in differential scanning calorimetry (DSC) experiments[33]. In the present study, a notable increase in storage and loss modulus at 1 Hz was observed in alginate/poloxamer gels with only 0.1 wt% sepiolite. However, when the sepiolite concentration exceeds 0.3 wt%, the reinforcement effect diminishes. This reduction is attributed to clustering of sepiolite particles within the matrix, which inhibits effective load transfer between the sepiolite and the alginate/poloxamer matrix, thereby compromising the mechanical performance. A similar trend is observed in the dynamic and loss moduli identified using the vibration transmissibility testing, with the effect being more pronounced at high temperatures. For instance, the 0.3 wt% sepiolite gels increased their loss modulus from 0.27 MPa at 30°C to 0.7 MPa at 60°C. Comparable trends are observed at lower sepiolite concentrations (0.1 wt%) and when 1 wt% cactus fibres are incorporated, with the loss modulus increasing from approximately 0.4 MPa at 30°C to 0.55 MPa at 60°C. Sepiolite particles possess a very high specific surface area with a negative surface potential[34]. Accordingly, direct interactions with the carboxylic acid groups on the alginate chains are unlikely; however, accumulation of the $Ca_{2}+$ ions via double layer formation may promote adhesion of the alginate chains to the particle surfaces. In summary, the larger values of storage and loss moduli at temperatures above 30°C for the sepiolite-based gels indicated the beneficial effect of sepiolite in bonding with the alginate calcium and improving the structural integrity of the gels at different temperatures.

4.2 Role of the cactus fibres

The incorporation of 1 wt% of cactus fibres increased the loss modulus of the alginate/poloxamer gels by approximately 33%. This effect appeared to be temperature-insensitive, although a slightly higher loss modulus was observed at 60°C. Increasing the weight fractions of cactus fibres enhanced load transfer between the reinforcements and the alginate/poloxamer matrix across all formulations tested in this study (ranging from 0.5 wt% to 2.0 wt%). Notably, no evidence of fibre clustering was observed with increasing cactus weight fraction, particularly in the vibration transmissibility tests. These findings supported the stiffening effect of low concentrations of cactus fibres, even in comparison to other natural fibres such as flax and bamboo[18]. Previous studies have attributed this behaviour to the multiscale fractal nature of cactus fibres, which induces slip-stick friction effects during dynamic fatigue, thereby increasing energy dissipation of composites[16]. The behaviour of the loss modulus (quasi-static and dynamic) also followed a similar trend here, confirming the results obtained from the vibration transmissibility tests carried out on the first generation of alginate/poloxamer/cactus fibre gels with a reinforcement level between 1.25 wt% to 5.0 wt%[19].

# 5. Surrogate modelling

The data from the design of experiments for the DMA and the vibration transmissibility tests were used to construct surface response metamodels (SRM)[35]. These SRMs are intended to support future optimisation of the manufacturing process using machine learning or AI-based techniques, by allowing prediction of outcomes based on varying parameters values specifically the weight fractions of sepiolite and cactus fibres The SRMs were developed using fifth- order polynomial fittings at the three different temperatures (30°C, 45°C and 60°C), under 20 %RH and 1 Hz of excitation. The coefficients of determination ($R^2$), for the polynomial fits ranged from 0.88 and 0.96, indicating a high level of agreement between the models and experimental data. The coefficients of the fittings for the DMA and the vibration transmissibility tests are provided in Appendix A and Appendix B, respectively. Figure A1 presents the SRMs derived from the data obtained from the DMA tests. These response surfaces indicate that the dispersion of cactus fibres contributes most significantly to variations in loss and storage moduli at a given temperature. In contrast, sepiolite is more effective in enhancing the viscoelastic response at elevated temperature. The most pronounced variations in tan δ are observed at intermediate weight fractions of sepiolite and cactus fibres. Figure A2 illustrates the SRMs extracted from the vibration transmissibility test data. The loss moduli in this dataset range of 0.15 MPa to 0.35 MPa, with the highest energy dissipation occurring in formulations containing at least 2wt% cactus fibres. Conversely, the lowest values for both dynamic moduli and resonant frequencies are associated with sepiolite weight fraction exceeding 0.3 wt%, even when the maximum loading of cactus fibres is used.

# Conclusions

The composite hydrogel materials developed through the Design of Experiments (DoE) approach in this study demonstrate an increase in quasi-static and storage modulus properties compared to the baseline alginate/poloxamer formulation. Storage modulus values were observed to be up to twice those of the undoped hydrogel, even with highly dilute dispersions of cactus fibre and sepiolite. Sepiolite reinforcements less than 0.3 wt% were effective in stabilising and stiffening the gel, particularly at elevated temperature. In contrast, cactus fibres provided a consistent increase in stiffness and damping with increasing weight fractions. These trends were confirmed though both Dynamic Mechanical Analysis (DMA) and vibration transmissibility testing. Notably, high values of loss factors (over ~0.4) during vibration testing were recorded in hydrogels with 0.1 wt% sepiolite and up to 2 wt% of cactus fibres. A formulation comprising 5 wt% alginate, 5 wt% poloxamer, 0.1 wt% sepiolite and 2 wt% cactus fibre emerged as the optimal configuration, balancing quasi-static stiffness with and dynamic performance.

Compared to previous generations of alginate/poloxamer hydrogels[30], the composites developed in this work, particularly those with 0.1 wt% sepiolite and 2 wt% cactus fibres, demonstrated up to an an order of magnitude increase (0.93 MPa) within the 80 Hz-140 Hz frequency range, relative to their static modulus. This is significant as the damping performance of polymers, especially fossil-based ones, is typically suboptimal at low frequencies[4]. Furthermore, vibration transmissibility tests revealed sensitivity of both loss factors and dynamic moduli to increasing base acceleration (up to 2.5 g[19]), suggesting possible granular-type fluidisation effects. Similar behaviours have been reported in jammed granular hydrogels systems for bioprinting applications[36]. The composite hydrogel (alginate/poloxamer/sepiolite/cactus fibre) also exhibited an average loss modulus around 0.32 MPa; a value that is comparable or exceeding those of conventional damping materials, such as commercial open cell polyurethane foams[28]. Remarkably these improvements were achieved using extremely dilute fibre loadings, matching the relative stiffening effects reported for alginate/PVA-based hydrogels that typically require 10 wt% reinforcement[10]. It is worth noting that the viscoelastic properties in this study were characterised under environmental conditions (e.g. relative humidity) that tends to dehydrate and degrade performance. Thus, further improved results may be expected under more controlled or milder testing conditions.

## Acknowledgements

This work has been funded by the UK Defence Science and Technology Laboratory (Grant ID: DSTL0000024791), the Office of Naval Research Global (Grant ID: N62909-201-2061), and the ERC H2020 AdG NEUROMETA project (01020715). The authors would also like to express their gratitude to Professor Petra C. Oyston, Mr Matthew D Eagling (Dstl), Dr Claire Lonsdale (Dstl), Dr Dave Hallam (Dstl), Maj. Anthony Kirkham (Dstl), Maj. James Barraclough (Dstl) and Dr Scott A. Walper (ONR G) for their support and insightful discussions throughout the project. The authors also thank Judith Mantell of the Wolfson Bioimaging Facility for their assistance in capturing the Cryo-SEM micrographs presented in this work. JPKA acknowledges funding from a UKRI Future Leaders Fellowship (MR/V024965/1).

# Appendix A – Surface response from the DMA tests

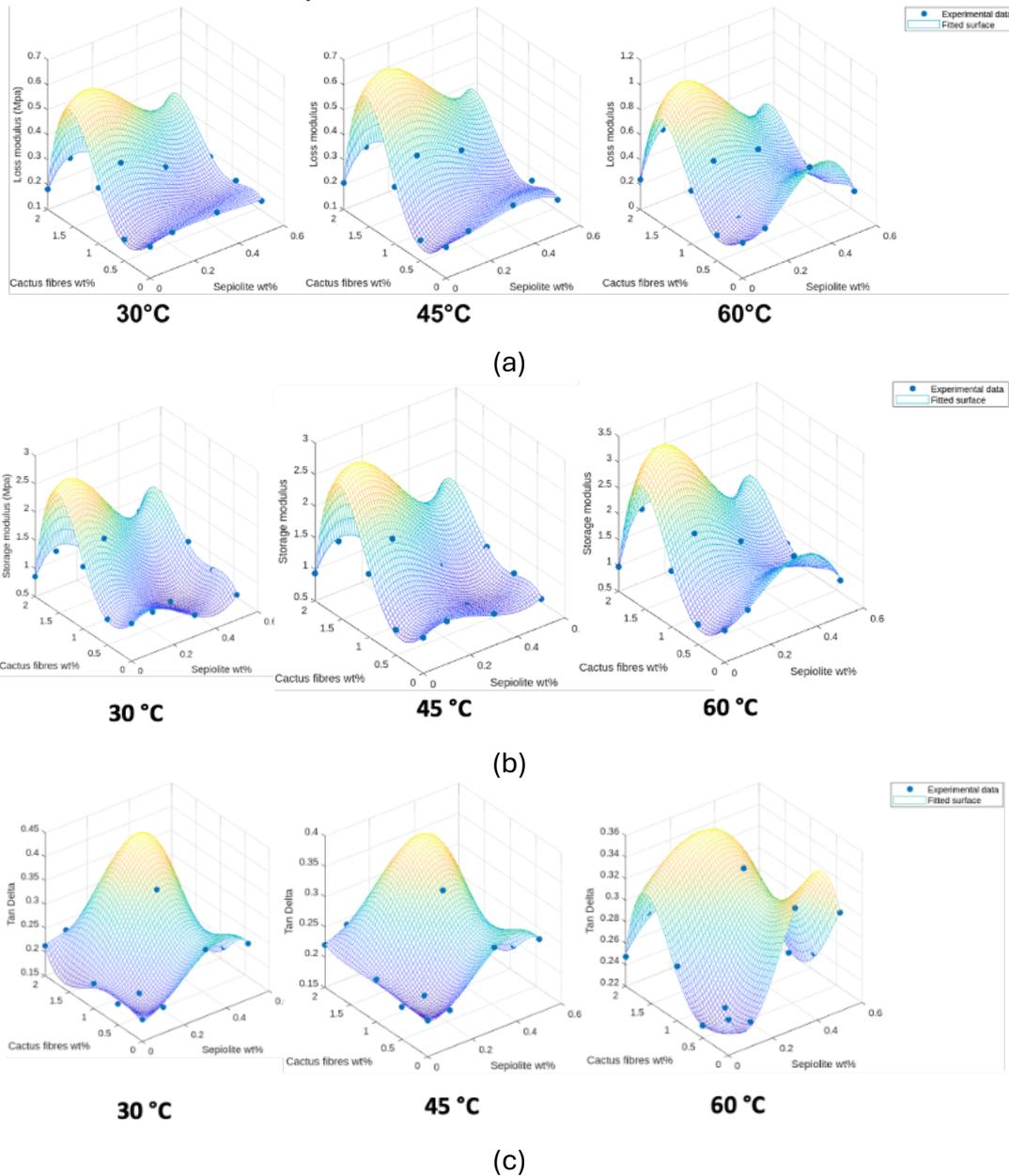

(a)

(b)

(c)

*Figure A1. Surfaces responses at different temperatures for (a) loss moduli, (b) storage moduli and (c) tan$\delta$.*

| Equation term | | Coefficient | Loss modulus (Mpa) R^2=0.9665 | Storage modulus (Mpa) R^2=0.9201 | Tan δ R^2=0.9366 |
|---|---|---|---|---|---|
| x1=Sepiolite wt% | x2=Cactus fibres wt% | | | | |
| | | a0 | 0.232 | 1.1428 | 0.1907 |
| x2 | | a1 | -0.2536 | -0.9659 | 0.0452 |
| x2 ^2 | | a2 | 0 | 0 | 0 |
| x2 ^3 | | a3 | 0.6667 | 2.8487 | -0.0538 |
| x2 ^4 | | a4 | -0.3051 | -1.3239 | 0 |
| x2 ^5 | | a5 | 0 | 0 | 0.0114 |
| x1 | | a6 | 0.228 | 3.31 | 0.0956 |
| x1 * x2 | | a7 | 1.6098 | 4.8187 | -3.5844 |
| x1 * x2 ^2 | | a8 | -1.6846 | -8.3329 | 3.7688 |
| x1 * x2 ^3 | | a9 | 0.6388 | 3.6014 | 0 |
| x1 * x2 ^4 | | a10 | 0 | 0 | -0.4949 |
| x1 ^2 | | a11 | -0.7646 | -24.216 | 1.6046 |
| x1 ^2 * x2 | | a12 | -4.3118 | -6.8573 | 13.6485 |
| x1 ^2 * x2 ^2 | | a13 | -0.0364 | -2.6024 | -13.4238 |
| x1 ^2 * x2 ^3 | | a14 | 0 | 0 | 3.144 |
| x1 ^3 | | a15 | 0.6775 | 33.593 | -2.9619 |
| x1 ^3 * x2 | | a16 | 5.8941 | 16.6323 | -12.8924 |
| x1 ^3 * x2 ^2 | | a17 | 0 | 0 | 6.9039 |

*Table A1.1. Coefficients for the storage, loss moduli and tan δ of the DMA tests at 45°C*

| Equation term | Coefficient | Loss modulus (Mpa) R^2=0.9351 | Storage modulus (Mpa) R^2=0.9525 | Tan δ R^2=0.9538 |
|---|---|---|---|---|
| x1=Sepiolite wt%    x2=Cactus fibres wt% | | | | |
| | a0 | 0.2352 | 1.0785 | 0.207 |
| x2 | a1 | -0.3089 | -1.0883 | 0.0126 |
| x2 ^2 | a2 | 0 | 0 | 0 |
| x2 ^3 | a3 | 0.7707 | 2.9645 | -0.0073 |
| x2 ^4 | a4 | -0.3487 | -1.3568 | 0 |
| x2 ^5 | a5 | 0 | 0 | 0.0015 |
| x1 | a6 | -0.0175 | 2.2946 | -0.0584 |
| x1 * x2 | a7 | 2.4919 | 7.4902 | -2.2066 |
| x1 * x2 ^2 | a8 | -2.0973 | -8.5185 | 2.3472 |
| x1 * x2 ^3 | a9 | 0.688 | 3.1685 | 0 |
| x1 * x2 ^4 | a10 | 0 | 0 | -0.2968 |
| x1 ^2 | a11 | 1.2099 | -14.3866 | 2.0191 |
| x1 ^2 * x2 | a12 | -6.9773 | -16.9131 | 8.7806 |
| x1 ^2 * x2 ^2 | a13 | 0.4964 | -0.0482 | -8.7984 |
| x1 ^2 * x2 ^3 | a14 | 0 | 0 | 1.9502 |
| x1 ^3 | a15 | -2.2601 | 18.6535 | -3.2792 |
| x1 ^3 * x2 | a16 | 7.9233 | 23.1414 | -8.7815 |
| x1 ^3 * x2 ^2 | a17 | 0 | 0 | 5.0332 |

*Table A1.2. Coefficients for the storage, loss moduli and tan δ of the DMA tests at 60°C*

| Equation term | Coefficient | Loss modulus (Mpa) R^2=0.9655 | Storage modulus (Mpa) R^2=0.9679 | Tan δ R^2=0.9655 |
|---|---|---|---|---|
| x1=Sepiolite wt%    x2=Cactus fibres wt% | | | | |
| | a0 | 0.2922 | 1.135 | 0.2562 |
| x2 | a1 | -0.4458 | -1.3099 | -0.0785 |
| x2 ^2 | a2 | 0 | 0 | 0 |
| x2 ^3 | a3 | 1.1167 | 3.4154 | 0.1543 |
| x2 ^4 | a4 | -0.5062 | -1.5561 | -0.0677 |
| x2 ^5 | a5 | 0 | 0 | |
| x1 | a6 | -0.6846 | 0.8755 | -0.6084 |
| x1 * x2 | a7 | 3.7714 | 9.0364 | 0.9461 |
| x1 * x2 ^2 | a8 | -2.1089 | -6.5741 | -0.3795 |
| x1 * x2 ^3 | a9 | 0.9677 | 3.249 | 0.0685 |
| x1 * x2 ^4 | a10 | 0 | 0 | 0 |
| x1 ^2 | a11 | 13.2573 | 17.7026 | 5.0561 |
| x1 ^2 * x2 | a12 | -20.5686 | -49.2297 | -3.9291 |
| x1 ^2 * x2 ^2 | a13 | -0.7943 | -3.9052 | 0.5038 |
| x1 ^2 * x2 ^3 | a14 | 0 | 0 | 0 |
| x1 ^3 | a15 | -23.2146 | -37.9938 | -7.2317 |
| x1 ^3 * x2 | a16 | 30.5701 | 79.2606 | 3.726 |
| x1 ^3 * x2 ^2 | a17 | 0 | 0 | 0 |

# Appendix B - Coefficients of the surface response from the vibration transmissibility tests

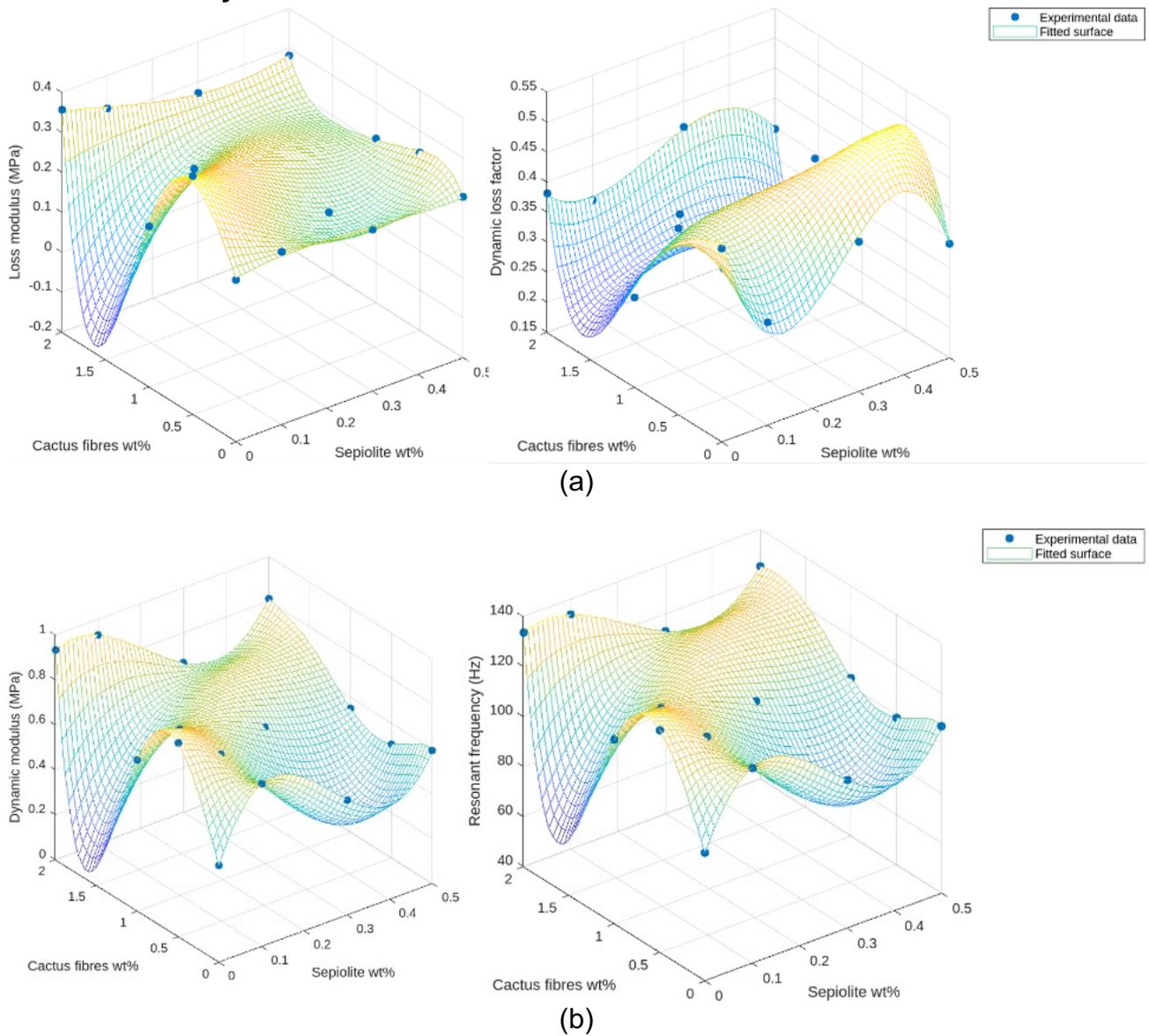

*Figure A2. Surface response models for (a) loss modulus and dynamic loss factors, and (b) dynamic modulus and resonant frequencies for the data related to the composite hydrogels measured using the vibration transmissibility tests.*

*Table A3. Coefficients of the SRMs for the dynamic, loss moduli and resonant frequencies*

| Equation term | Coefficient | Dynamic modulus (Mpa) R^2=0.9999 | Dynamic loss factor R^2=0.8851 | Loss modulus (Mpa) R^2=0.9544 | Resonant frequency (Hz) R^2=0.9998 |
|---|---|---|---|---|---|
| x1=Sepiolite wt%   x2=Cactus fibres wt% | | | | | |
| | a0 | 0.4329 | 0.4816 | 0.2103 | 91.1534 |
| x2 | a1 | 1.1151 | -0.0571 | 0.4851 | 98.191 |
| x2 ^2 | a2 | 0 | 0 | 0 | 0 |
| x2 ^3 | a3 | -1.0931 | -0.1428 | -0.6078 | -94.6437 |
| x2 ^4 | a4 | 0 | 0 | 0 | 0 |
| x2 ^5 | a5 | 0.2192 | 0.036 | 0.1261 | 18.8561 |
| x1 | a6 | 5.2252 | -3.1219 | 0.2016 | 487.1795 |
| x1 * x2 | a7 | -20.4426 | 6.1747 | -5.4617 | -1863.8346 |
| x1 * x2 ^2 | a8 | 19.094 | -2.0875 | 7.7134 | 1734.7081 |
| x1 * x2 ^3 | a9 | 0 | 0 | 0 | 0 |
| x1 * x2 ^4 | a10 | -2.4843 | -0.0994 | -1.2776 | -226.2982 |
| x1 ^2 | a11 | -26.5872 | 15.5565 | -0.9501 | -2467.8144 |
| x1 ^2 * x2 | a12 | 56.4367 | -29.1862 | 5.418 | 5109.1376 |
| x1 ^2 * x2 ^2 | a13 | -52.1447 | 9.1966 | -17.6422 | -4707.8775 |
| x1 ^2 * x2 ^3 | a14 | 13.803 | 1.3387 | 7.549 | 1268.4255 |
| x1 ^3 | a15 | 33.5585 | -19.7863 | 1.0076 | 3112.9735 |
| x1 ^3 * x2 | a16 | -38.9232 | 39.4368 | 10.8457 | -3440.0568 |
| x1 ^3 * x2 ^2 | a17 | 15.797 | -16.4585 | -5.3613 | 1313.447 |